\documentclass{sig-alternate-hotpets15}
\usepackage[final]{trackchanges}
\graphicspath{{./images/}}
\usepackage{listings}
\usepackage{url}
\usepackage{amsmath}
\usepackage[usenames,dvipsnames,svgnames,table]{xcolor}
\usepackage{epstopdf}
\date{}
\definecolor{MyDarkBlue}{rgb}{0,0.08,0.45}
\definecolor{lightgray}{rgb}{0.9,0.9,0.9}
\definecolor{lightlightgray}{rgb}{0.95,0.95,0.95}
\begin{document}

\title{ Immutable Views}
\subtitle{Access control (to your information) for masses}
%
%
%
%
%

\numberofauthors{1} 
%
\author{
%
%
\alignauthor
Yan Shvartzshnaider\\
       \affaddr{Princeton University}\\
       \email{yansh@princeton.edu}
}
\date{30 July 1999}

\maketitle
\begin{abstract}
There are a lot of on going efforts in the research community as well as industry around providing privacy-preserving and secure storage for personal data. Although, over time it has adopted many tag lines such as Personal Information Hub~\cite{website:ucn}, personal container~\cite{mortier2010personal}, DataBox~\cite{haddadi2015personal}, Personal Data Store (PDS)~\cite{de2014openpds} and many others, these are essentially reincarnations of a  simple idea: provide a secure  way and place for users to store their information and allow them to provision who has access to that information. 

In this paper, we would like to discuss a way to facilitate access control mechanism (AC) in the various ``personal cloud" proposals. 
\end{abstract}

%
%

\section{Introduction}

Personal information storage is the something we seem to take for granted. Today, we choose to store our data in  varied ``clouds" because we love the simplicity and convenience these services provide. However, this flourishing relationship relies on the implicit trust between users and service providers that users' privacy will remain intact.  While often we choose to believe that (major) service providers are the good guys\footnote{This despite explicit statements by some of the services that their business models are mainly depend on going through our private information.}, recent cases, where this trust was broken, are forcing us to realize that things are not as rosy as they might seem~\cite{ion2011cloud}. We gradually become aware of the fact that our information is being exploited by the big cooperations for a relatively low price---free service.  

\emph{How to ensure the our information is safe and how to (re)gain control?} These questions are currently preoccupy many researchers. The answers are not trivial and will probably require a clean-slate approach in how we build and manage our information systems.  In addition to exploring new business models~\cite{ng2013making} and incentives to change the general behavior of everyone involved. 

Recent efforts~\cite{website:ucn, mortier2010personal, haddadi2015personal, de2014openpds} look at facilitating ``personal clouds" to store users' data. While many seem to agree that such approach can offer a viable alternative, questions on how such systems should be designed and implemented remain open. 

On one extreme, works like~\cite{boshevski2014shawn} offer a completely decentralised cloud using P2P-based approach: storage is crowdsourced from a pool of participating users. This approach requires strong incentive schemes to reach critical mass. Alternatively, we can also use various cryptographic approaches to help us secure our information. 
But one can never be sure, so other approaches~\cite{bessani2013depsky} employ secret sharing techniques to split and spread content's chunks across different third party clouds. Although, it prevents any single third party from obtaining the whole content, this approach limits the access control to share all or nothing options.

However, what if the goal is to share parts of the information? Users might want to share some of their information with a third party to benefit from its services.  

In this paper we propose an approach for provisioning third party access to the bits released by the user. Our approach is based on snapshots, we call them ``immutable views". The views are created by running a query against a personal information space. The result is an immutable view containing only the information that the requester (and only him) can ``see".  The requester only deals with views, raw information is never retrieved. This means that immutable views  can be shared and stored by a third party without compromising user's privacy---since only the authorized requester can query it.  

We believe, the ideas described in this paper can be leveraged by many of the ongoing efforts, however, our implementation aligns well with the platform offered by the European User-Centric Networking project~\cite{website:ucn}. In particular, our system uses Irmin~\cite{website:irmin}, a git-like distributed store and Mirage OS~\cite{website:mirage}. Nonetheless, our current implementation~\cite{website:moanaml} is  programmed to support other backends.
\section{Information sharing}\label{lbl:protocol}

Our AC model is based on Moana~\cite{shvartzshnaider2014moana} service abstraction. Moana provides a graph-based service abstraction and information model. The Moana service model exposes two functions \verb+ADD+ and \verb+MAP+. The \verb+ADD+ function annotates the graph, while the \verb+MAP+ function is a standing graph query on the global information space.

\begin{figure}
      \includegraphics[width=0.49\textwidth]{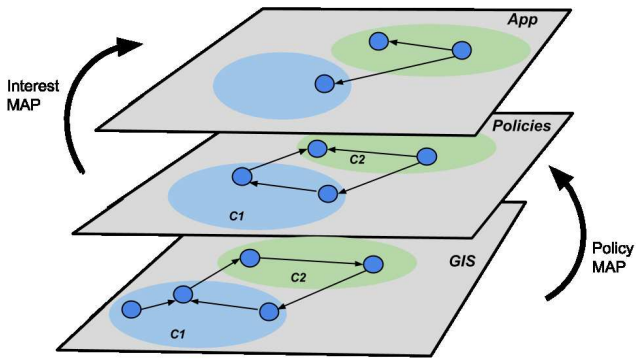}
    \caption{Three layers of AC control}
    \label{fig:layers}
\end{figure}

\textbf{AC model.}  As depicted by Figure~\ref{fig:layers}, the AC model comprises three conceptual views. The first view from the bottom---global information space (GIS)---contains all the facts and made assertions. Note that assertions also include the policies used to determine what gets through to the second view. The second view, is a result of a query, referred by us as policy \verb+MAP+, onto the first view  for all the information the requester can access. Finally, the last view is specific to the requester's query, we refer to it as an interest \verb+MAP+.  

In other words, any request triggers a chain of \verb+MAP+ operations that narrows down the information scope accessible to the  requester.

\textbf{Protocol.}  In layman's terms, we propose git inspired protocol for sharing information which enforces the above AC model. To store information, the user will \emph{init} a master information repository (IR), which will store all of its information. This essentially creates the first view, which the owner can populate with relevant AC polices. A third party service that would want to access it for their purposes will need to \emph{clone} it into a local view. ``Clone" operations rely on the policy \verb+MAPs+ in the AC model.  This ensures that only permitted information is cloned. The third party service is then left with a ``copy"  they can use for querying (using interest \verb+MAPs+) and running their algorithms. It is worth noting again that the query doesn't return any raw information, rather an immutable view wrapped in a sandbox container that can only be queried according to the specified AC policies. This is similar to SafeAnswers proposal in \cite{de2014openpds}, which only returns \textit{answers}  to the requester, not raw metadata. For view updates, the \verb+ADD+ function is used, which creates a new immutable instance of a view. An updated view from the master repository will be \emph{pulled} and new view instances created by third parties will be \emph{pushed} to the master repository, following respective \emph{merge} of the views.

\textbf{Social network example.} To illustrate intuitively how we can use this protocol, let's consider how a simple social network can be implicitly built on top of it.  Our social network follows a simple model: users have followers, who are permitted to get updates  e.g., on new photos, status updates, events, etc. They can also comment on users' content. Users publish new information such as events, photos and status updates to their personal IR. The repository, as previously mentioned, contains and enforces the AC policies describing who can see what. Users' followers, those who  wish to obtain his content will clone the repository. Consequently, similar to git, the followers will be able to pull updates from the master IR and also push (if permitted) their comments, as a new view instance, to be merged with the master view.


\textbf{Implementation.}  For our system design we are looking at leveraging some recent work into unikernels~\cite{madhavapeddy2013unikernels} based on Mirage OS~\cite{website:mirage}.  In particular, we envision IRs served using unikernels. Unikernels are virtual machines that run specialised Operating System (OS) to support individual software components. Mirage OS provides with the needed flexibility in deployment and management. In particular, thanks to Mirage OS, unikernels can be compiled and run against any environment such as a personal computer, custom open-source hardware and cloud. Most importantly, unikernels can be programmed with internal logic to ensure that users' policies will be respected. 
Once a requester authenticates against the IR service, hosted on a unikernel, another unikernel is spun off containing information subset (in an immutable view) to which the requester has access.  The requester will essentially interact with IR service specially ``designed" for him.

{ \bibliographystyle{acm}
\bibliography{references}}
\end{document}